\documentclass{INTERSPEECH2023}

\interspeechcameraready 

\title{Continual Learning for End-to-End ASR by Averaging Domain Experts}
\name{Peter Plantinga, Jaekwon Yoo, Chandra Dhir}
\address{JP Morgan Chase \& Co., USA}
\email{peter.plantinga@chase.com, jaekwon.yoo@jpmchase.com, chandra.dhir@jpmchase.com}

\begin{document}

\maketitle
 
\begin{abstract}
Continual learning for end-to-end automatic speech recognition has to contend with a number of difficulties. Fine-tuning strategies tend to lose performance on data already seen, a process known as catastrophic forgetting. On the other hand, strategies that freeze parameters and append tunable parameters must maintain multiple models. We suggest a strategy that maintains only a single model for inference and avoids catastrophic forgetting.

Our experiments show that a simple linear interpolation of several models’ parameters, each fine-tuned from the same generalist model, results in a single model that performs well on all tested data. For our experiments we selected two open-source end-to-end speech recognition models pre-trained on large datasets and fine-tuned them on 3 separate datasets: SGPISpeech, CORAAL, and DiPCo. The proposed average of domain experts model performs well on all tested data, and has almost no loss in performance on data from the domain of original training.

\end{abstract}

%
%
\section{Introduction}

Modern end-to-end automatic speech recognition (E2E-ASR) systems have achieved impressive results across a variety of data by training on massive datasets up to 700,000 hours~\cite{radford2022robust}. While these generalist models often perform surprisingly well on domains they have never seen in a zero-shot manner, for specific applications they can still benefit tremendously from fine-tuning on data from the target domain.

A typical strategy for fine-tuning E2E-ASR systems involves standard gradient descent updates to model parameters using data from the target domain~\cite{mirsamadi2017multi}. However, this strategy usually suffers reduced performance on data from the original domain, a process known as catastrophic forgetting~\cite{parisi2019continual}. While for some it may be possible to maintain different parameters for different domains this has the downside of adding complexity and taking up storage space, especially for large models. In addition, it may not be clear for all cases which domain a target sample falls into.

Some have sought to address this difficulty with special attention paid to certain parameters, either by freezing some parameters~\cite{Takashima2022UpdatingOE} or by adding loss regularization designed to reduce forgetting~\cite{Li2016LearningWF}. These techniques have met mixed success with mitigating forgetting; serial fine-tuning processes on new domains still often results in decreased performance on the original data. We address this limitation by parallelizing the fine-tuning process and averaging the parameters of the fine-tuned expert models.

Others have addressed this difficulty by freezing the entirety of the good-performing generalist model and adding domain-specific parameters~\cite{Rusu2016ProgressiveNN}. One popular technique along these lines is called Adapters~\cite{Wang2020KAdapterIK}, which involves freezing original model parameters and updated small modules inserted with a starting configuration which preserves the behavior of the original model. A weakness of this approach is that multiple sets of parameters must be maintained, and at inference a decision must be made about which set to use.

One final technique involves replaying data from the original domain 
\cite{Isele2018SelectiveER}. This approach can work well when the original data is available but is not always possible, especially for pretrained models where the original data is not publicly available.

To summarize the contributions of this work, we re-formulate the continual learning paradigm from many serial applications of fine-tuning to a single model into a parallel learning process whereby multiple fine-tuned domain expert models are averaged into a single good-performing model. We call this paradigm Average of Domain Experts (AoDE). 

%
%
\section{Related work}

\begin{figure*}[t]
  \centering
  \includegraphics[width=\linewidth]{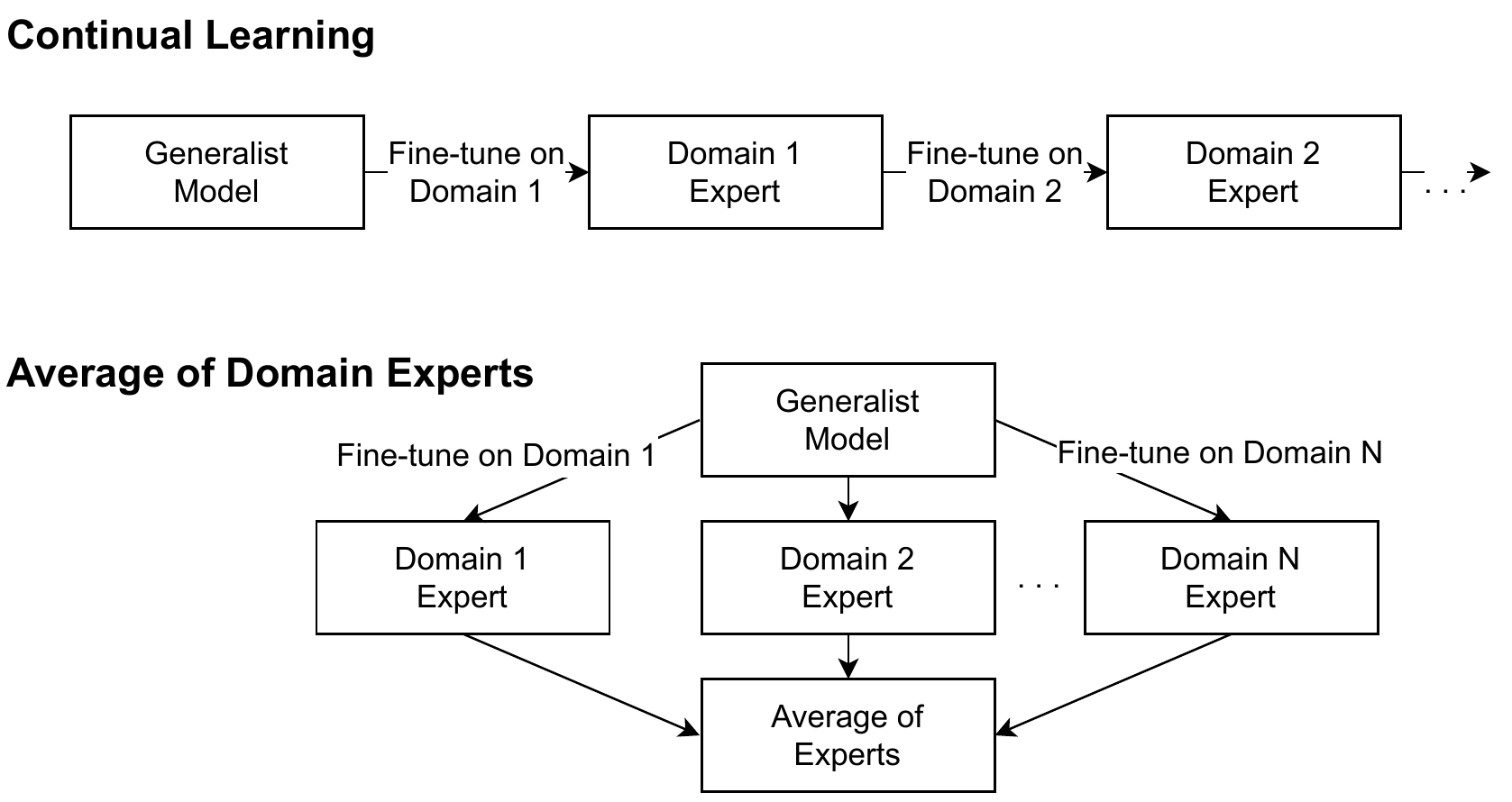}
  \caption{Our proposed update to the continual learning paradigm: instead of training sequentially on a variety of domains, fine-tune on each domain in parallel and then combine the results to get an average of domain experts model with no forgetting.}
  \label{fig:paradigm}
\end{figure*}

Model parameter averaging appears in a number of contexts but appears only rarely in the context of continual learning, given the emphasis of the field on serial fine-tuning on a sequence of new domains. We relate a few of these contexts here:

For distributed model training with limited connectivity, called federated learning, some researchers have found that averaging models can achieve a similar performance as training a single model on all data~\cite{kamp2019efficient, wang2020federated}. This strategy has the advantages of reducing communication overhead, as well as preserving data privacy by sharing only model parameters and not data samples. Dynamic approaches share parameters more or less frequently based on how rapidly performance deteriorates on out-of-domain data.

Another context that model averaging appears in is improving generalization of trained models. One example is Stochastic Weight Averaging~\cite{Izmailov2018AveragingWL} which finds better optima by collecting checkpoints throughout the training process and averaging them.

Other researchers have used model averaging to improve semi-supervised learning using a teacher model~\cite{tarvainen2017mean}. The authors found that better teacher models could be created by averaging the parameters of several teacher models, created by adding noise to the internal representations of student models. These averaged teacher models produce better training targets during semi-supervised training.

One last example is that model averaging is used to understand the dynamics of loss basins for neural networks~\cite{ainsworth2022git}. In order to understand how it is that different random initializations of ResNets end up achieving very similar performance after training, the authors suggest permuting the parameters in an isomorphic way so as to merge the distinct loss basins into a single loss basin. This is done by rearranging the parameters in one model to best match the parameters in an original model, and then merging the models.

All of this related work shows that model parameter averaging is a powerful technique that is under-used for the purpose of continual learning.

%
%
\section{Experiments}

We ran experiments on two large end-to-end speech recognition models pretrained on large sets with diverse data. Our experiments involve fine-tuning on three separate public datasets with diverse qualities. Details on models, datasets, fine-tuning procedure, and evaluation procedure are explained in the following sections.

\subsection{Pretrained Models}
\label{ssec:pretrained}

We demonstrate the generality of our results by using unrelated pre-trained models with differing architecture, training loss, data, sponsoring organization etc. The two pretrained models we used in our experiments are the NeMo Conformer CTC Large model~\cite{kuchaiev2019nemo} and the OpenAI Whisper Small.en model~\cite{radford2022robust}.

The NeMo English Conformer CTC Large model is trained with Connectionist Temporal Classification (CTC) loss and consists of a small downsampling layer followed by 18 convolution + self-attention blocks and a final output layer. The total number of parameters is 121M.

The tokenizer vocabulary includes 128 subword tokens; all tokens include only lowercase letters, apostraphes, and spaces. The acoustic model is trained on NeMo ASRset which consists of roughly 25,000 hours of audio from a variety of sources, most of which are publicly available. We use version 1.10 of the model, the most recent version at the time of submission.

The Whisper Small.en model~\cite{radford2022robust} is trained using standard sequence-to-sequence cross-entropy loss and consists of two major sub-models, an encoder and a decoder. The encoder consists of a small downsampling layer followed by 11 self-attention blocks and the decoder consists of 11 multi-headed attention blocks and an output layer. The total number of parameters is 241M.

This model uses an English-only tokenizer with the same 50k-token vocabulary as GPT-2. The set of characters present in these tokens is much larger than for NeMo Conformer, including upper and lower case as well as punctuation. The training data consists of roughly 500,000 hours of English-only data present in the OpenAI speech data.

\subsection{Datasets}

For our experiments we used three public datasets with a large variety of qualities. For our first dataset, we used SPGISpeech \cite{oneill2021spgispeech}, which consists of 5000 hours of high-quality recordings of earnings calls. These recordings are well-transcribed and are difficult for a generalist model only on account of a large vocabulary of financial terms that are unlikely to appear elsewhere. Since the original dataset contains punctuation and numbers but the NeMo Conformer tokenizer cannot encode these values, we pass the transcript through NeMo normalization~\cite{Zhang2021NeMoIT} when training and evaluating NeMo Conformer. To speed up evaluation, we use a random subset of 2000 samples (roughly same size as LibriSpeech test sets) for a test set, and find the resulting performance differs by less than 3\% relative in all measured cases. In addition, while training the Whisper Small.en model we found that our techniques were still not sufficient to prevent some catastrophic forgetting when the entire training set was used. Instead, we select a random subset of about 10\% of the data for training.

The second dataset we used was the CORAAL dataset~\cite{kendall2021}, a conversational dataset between folks whose primary sociolect is African American Vernacular English (AAVE). The data was recorded in six separate locations and over the course of ten years (with one exception). In total, there are more than 150 interviews at a length surpassing 140 hours of audio. We split the data by separating 5 speakers for each of validation and test sets, amounting to roughly 5 hours each. Generalist models have difficulty with the conversational nature of the data and the different grammars of the sociolect. We divided the audio into segments based on the provided timings in the transcript, with total length not exceeding 30s, in order to match the expected input length for Whisper models.

Finally, we experimented with the DiPCo dataset~\cite{Segbroeck2019}, a small dataset of conversation in a dinner party scenario. These data were the most challenging, involving the most speakers and varied acoustic conditions. The length of the audio available is 2.7 hours for development and 3.4 hours for test. In a process similar to the one used on CORAAL data, we divided the audio into segments not exceeding 30s in length. The conversations were recorded from a number of devices, for the sake of simplicity we take the sum of close-talking microphones as the audio signal.

\subsection{Fine-tuning Procedure}

\begin{table}[t]
  \caption{WER performance comparison for Whisper Small.en fine-tuning with several values of LLRD.}
  \label{tab:llrd}
  \centering
  \begin{tabular}{cccc}
    \toprule
    Model     & CORAAL  & LibriSpeech & LibriSpeech \\
    procedure & test set & test-clean & test-other \\
    \midrule
    Pretrained    & 18.8 & 3.35 & 7.55 \\
    LLRD=1.0      & 14.4 & 5.86 & 11.8 \\
    LLRD=0.9      & 12.4 & 3.82 & 9.01 \\
    LLRD=0.8      & 13.2 & 3.32 & 8.31 \\
    \bottomrule
  \end{tabular}
  
\end{table}

Nearly all SGD-based fine-tuning procedures that sequentially access data will inevitably lose some performance on the original domain. In addition, the choice of which order to fine-tune on domains can have a significant effect on the outcome (see discussion of Table~\ref{tab:nemo} later). To address all of these limitations and produce a single well-performing generalist model with no loss of performance, we propose a new continual learning paradigm that fine-tunes on each domain in parallel, and then averages the resulting expert models. See Figure~\ref{fig:paradigm} for a graphical depiction of the proposed Average of Domain Experts approach.

We begin by reproducing state-of-the-art continual learning techniques such as layer-wise learning rate decay (LLRD)~\cite{Zhang2020RevisitingFB} and slanted triangular learning rates (STLR)~\cite{Howard2018UniversalLM}. These techniques definitely help with better learning and less forgetting, as shown in Table~\ref{tab:llrd}. LLRD is applied by assigning the highest learning rate to the highest encoder layer and decaying the learning rate of each lower layer by a constant factor, usually 0.9. The learning rate of the lowest encoder layer is applied to any layers not in the encoder (decoder, output, embedding, etc.), from the inspiration of~\cite{Takashima2022UpdatingOE}. Our learning rate schedule, STLR, peaks at roughly 10-20\% of the total training time.

Freezing non-encoder layers, as suggested by~\cite{Takashima2022UpdatingOE} is roughly equivalent to reducing the overall learning rate in the proposed scheme, as shown in Table~\ref{tab:llrd_frozen}. If one compares the frozen layers row with the following row, they achieve very similar results across the board. We also found no benefit to LLRD by adding a loss against the predictions of the original model, a technique called Learning without Forgetting (LwF)~\cite{Li2016LearningWF}.

Once the fine-tuning process is done producing domain expert models, we compute the average of experts in a straightforward way: a simple linear interpolation of corresponding model parameters with equal weighting on every model. This is sufficient for good results, and works well with other techniques for reduced forgetting, such as LLRD.

All experiments are conducted with the SpeechBrain toolkit~\cite{speechbrain} on a single machine with four 24GB A10 GPUs. Batch size was maximized for the available space (5 per GPU for Whisper, dynamic batching at about 15 per batch for the longest samples for NeMo Conformer). We used Adam optimizer with default hyperparameters other than learning rate. Learning rate and LLRD rate were the sole optimized hyperparameters. 

\begin{table}[t]
  \caption{WER performance comparison between freezing non-encoder layers as proposed by \cite{Takashima2022UpdatingOE} and the LLRD approach, using NeMo Conformer CTC model.}
  \label{tab:llrd_frozen}
  \centering
  \begin{tabular}{cccc}
    \toprule
    Model & SPGI & LibriSpeech & LibriSpeech \\
    procedure & test set & test-clean & test-other \\
    \midrule
    Pretrained Conformer   & 5.47 & 2.15 & 4.48 \\
    Frozen layers, lr=3e-4 & 2.73 & 2.74 & 5.84 \\
    LLRD=0.9, lr=1e-4      & 2.74 & 2.77 & 5.94 \\
    LLRD=0.9, lr=3e-4      & 2.63 & 2.98 & 6.52 \\
    \bottomrule
  \end{tabular}
\end{table}

\subsection{Evaluation}

\begin{table*}[t]
  \caption{NeMo Conformer WER results on five test sets and the geometric mean of the five scores.}
  \label{tab:nemo}
  \centering
  \begin{tabular}{lccccccc}
    \toprule
    Model procedure & SPGI test & CORAAL test & DiPCo test & LS test-clean & LS test-other & Geometric mean \\
    \midrule
    Pretrained model     & 5.47 & 28.6 & 70.7 & 2.15 & 4.48 & 10.1 \\
    Fine-tuned on SPGI   & 2.63 & 23.9 & 70.5 & 2.98 & 6.52 & 9.71 \\
    Fine-tuned on CORAAL & 5.99 & 15.6 & 45.4 & 3.06 & 6.70 & 9.72 \\
    Fine-tuned on DiPCo  & 7.01 & 25.3 & 47.3 & 3.15 & 6.95 & 11.3 \\
    SPGI $\rightarrow$ CORAAL & 3.52 & 16.7 & 44.8 & 3.10 & 6.91 & 8.92 \\
    SPGI $\rightarrow$ CORAAL $\rightarrow$ DiPCo & 4.07 & 19.7 & 46.3 & 3.48 & 7.70 & 10.0 \\
    SPGI $\rightarrow$ DiPCo $\rightarrow$ CORAAL & 3.57 & 16.4 & 44.1 & 3.09 & 6.85 & 8.86 \\
    Average of SPGI and CORAAL & 3.25 & 17.5 & 51.3 & 2.33 & 5.02 & 8.07 \\
    Average of Domain Experts & 3.04 & 18.2 & 45.2 & 2.18 & 4.86 & 7.67 \\
    \bottomrule
  \end{tabular}
  
\end{table*}

\begin{table*}[th]
  \caption{Whisper Small.en WER results on five test sets and the geometric mean of the five scores.}
  \label{tab:whisper}
  \centering
  \begin{tabular}{lccccccc}
    \toprule
    Model training procedure & SPGI test & CORAAL test & DiPCo test & LS test-clean & LS test-other & Geometric mean \\
    \midrule
    Pretrained model     & 4.94 & 18.8 & 48.5 & 3.35 & 7.55 & 10.3 \\
    Fine-tuned on SPGI   & 2.87 & 22.6 & 50.1 & 5.50 & 11.1 & 11.5 \\
    Fine-tuned on CORAAL & 4.58 & 12.4 & 44.3 & 3.82 & 9.01 & 9.72 \\
    Fine-tuned on DiPCo  & 4.31 & 18.2 & 44.0 & 3.26 & 7.63 & 9.70 \\
    SPGI $\rightarrow$ DiPCo $\rightarrow$ CORAAL & 4.35 & 13.1 & 43.3 & 3.35 & 8.23 & 9.26 \\
    Average of SPGI and CORAAL & 3.09 & 15.9 & 44.0 & 3.52 & 8.13 & 9.08 \\
    Average of Domain Experts & 3.41 & 15.7 & 43.0 & 3.39 & 7.74 & 9.04 \\
    \bottomrule
  \end{tabular}
  
\end{table*}

%

As noted in section~\ref{ssec:pretrained}, NeMo Conformer Large CTC and Whisper Small.en have different output character sets, and therefore we have different normalization processes for the texts before WER computation.

For the NeMo model evaluations, we first normalized the transcript according to the same process used during training target preparation: we used NeMo Text Normalizer to do FST-based conversion of numbers and some symbols to pure text (e.g. \$5~$\rightarrow$~five dollars and 5:00~$\rightarrow$~five o'clock). Then all punctuation was stripped and text lower-cased. On top of these normalizations we added a few new ones: we normalized contractions to the shortened form and removed hesitations (e.g. um, hmm). For decoding, we simply used greedy decoding with no language model.

For the Whisper model evaluations, we used the English text normalizer provided as part of the model code, which performed many of the same normalizations listed above, as well as some spelling normalizations. We also relied on Whisper model code to perform decoding. The code by default does greedy decoding and automatically handles certain failure cases. If compressing the predicted transcript surpasses a given compression ratio, indicating the autoregressive decoder is stuck in a loop, then the output is regenerated using a different sampling temperature. Similarly, if the average log probability over sampled tokens is below some threshold, the output is regenerated. We used the default values for all thresholds.

%
%
\section{Results}

Our experimental results are shown in Tables \ref{tab:nemo} and \ref{tab:whisper}, with results on tests sets from the three domains used for training, as well as LibriSpeech test-clean and test-other. The LibriSpeech datasets provide a measure of catastrophic forgetting, though in the case of Whisper the model was likely never trained on the LibriSpeech data. To summarize the results on each domain we report the geometric mean, in order to avoid a bias towards more difficult domains, given the wide spread of WERs.

In the results tables, fine-tuned models are all trained using the proposed procedure, including LLRD and STLR. After listing individually fine-tuned models, a list of sequentially tuned models are shown using an arrow to represent the progression of the tuning process. Finally, the proposed average of domain experts method is listed.

\subsection{NeMo Conformer}

The experimental results for NeMo Conformer can be seen in Table~\ref{tab:nemo}. The first point of note is that the averaged models have substantially lower geometric means relative to all other model training procedures. In large part this is driven by a reduction in catastrophic forgetting, with final test performances at or very near the performance of the original pretrained model.

In addition to the drastic reduction in catastrophic forgetting, the proposed model's improvement is within 10\% relative of the best performing model for each test set. Compared to the best-performing model on each domain, the averaged model is within 7.5\% relative on SGPI, within 9\% relative on SPGI, and within 2\% relative on DiPCo.

Another point of note in this table is the dramatic effect of the order of datasets for sequential fine-tuning; the order SPGI~$\rightarrow$ CORAAL~$\rightarrow$ DiPCo performs worse than SPGI~$\rightarrow$ DiPCo~$\rightarrow$ CORAAL in every category. This demonstrates some of the difficulties with sequential fine-tuning that would not be present in the average of domain experts approach.

\subsection{Whisper Small.en}

The experimental results for Whisper Small.en can be seen in Table~\ref{tab:whisper}. 
Again, the proposed average of experts achieves the lowest geometric mean, mainly due to reduced forgetting on LibriSpeech test-other and SPGI test.

We found the Whisper model more susceptible to catastrophic forgetting than the NeMo model; using the full SPGI training set resulted in a model that, when averaged with other models, produced WERs close to 100\%. As mentioned before, we ended up using 10\% of SPGI training data. We speculate that fine-tuning the model on the full set rearranges its parameters enough to fall into a separate loss basin. It might be possible to recover the original parameter arrangement by accounting for permutation invariances~\cite{ainsworth2022git}.

In addition, we used a more aggressive LLRD of 0.8 for the sequential fine-tuning, resulting in less forgetting (and slightly worse performance on each individual set). This likely contributes to the improved geometric mean score for sequential fine-tuning, however this model still demonstrates some forgetting on the first fine-tuning set (i.e. SPGI). While one could argue this provides a degree of control over what domains the model performs best on, we argue that increasing or decreasing the proportion of each model present in the final average gives more fine-grained control.

%
%
\section{Conclusions}

We were able to show that a simple update to the continual learning paradigm is able to dramatically reduce catastrophic forgetting in well-trained generalist end-to-end speech recognition models. This is done by a simple average of experts, proving a flexible and tunable approach to model development.

In the future, we hope to improve on this work with more sophisticated averaging techniques, taking into account permutation invariances, with techniques such as Git Re-Basin~\cite{ainsworth2022git} or Federated Matched Averaging (FedMA)~\cite{wang2020federated}.

\bibliographystyle{IEEEtran}
\bibliography{mybib}

\end{document}